\documentstyle[12pt]{article}
\input epsf

\begin{document}
\baselineskip=15pt
\newcommand{\x}{{\bf x}}
\newcommand{\y}{{\bf y}}
\newcommand{\z}{{\bf z}}
\newcommand{\bp}{{\bf p}}
\newcommand{\A}{{\bf A}}
\newcommand{\B}{{\bf B}}
\newcommand{\p}{\varphi}
\newcommand{\del}{\nabla}
\newcommand{\be}{\begin{equation}}
\newcommand{\ee}{\end{equation}}
\newcommand{\bq}{\begin{eqnarray}}
\newcommand{\eq}{\end{eqnarray}}
\newcommand{\ba}{\begin{eqnarray}}
\newcommand{\ea}{\end{eqnarray}}
\def\r{\nonumber\cr}
\def\hf{\textstyle{1\over2}}
\def\qr{\textstyle{1\over4}}
\def\Sc{Schr\"odinger\,}
\def\sc{Schr\"odinger\,}
\def\'{^\prime}
\def\>{\rangle}
\def\<{\langle}
\def\-{\rightarrow}
\def\dbd{\partial\over\partial}
\def\tr{{\rm tr}}

\begin{titlepage}
\vskip1in
\begin{center}
{\large Order $1/N^2$ test of the Maldacena conjecture II: the
full bulk one-loop contribution to the boundary Weyl anomaly  }
\end{center}
\vskip1in
\begin{center}
{\large Paul Mansfield$^a$, David Nolland$^b$ and Tatsuya Ueno$^b$

\vskip20pt

$^a$Department of Mathematical Sciences

University of Durham

South Road

Durham, DH1 3LE, England

{\it P.R.W.Mansfield@durham.ac.uk}

\vskip20pt

$^b$Department of Mathematical Sciences

University of Liverpool

Liverpool, L69 3BX, England

{\it nolland@liv.ac.uk}

{\it ueno@liv.ac.uk} }

\end{center}
\vskip1in
\begin{abstract}

\noindent We compute the complete bulk one-loop contribution to
the Weyl anomaly of the boundary theory for IIB Supergravity
compactified on $ AdS_5\times S^5$. The result, that $\delta {\cal
A}=(E+I)/\pi^2$, reproduces the subleading term in the exact
expression ${\cal A}=-(N^2-1)(E+I)/\pi^2$ for the Weyl anomaly of
${\cal N}=4$ Super-Yang-Mills theory, confirming the Maldacena
conjecture. The anomaly receives contributions from all multiplets
casting doubt on the possibility of describing the boundary theory
beyond leading order in $N$ by a consistent truncation to the
`massless' multiplet of IIB Supergravity.

\end{abstract}

\end{titlepage}


Henningson and Skenderis' beautiful computation \cite{Henningson}
of the Weyl anomaly of ${\cal N}=4$ $SU(N)$ Super-Yang-Mills
theory from five dimensional gravity is a remarkable test of the
Maldacena conjecture \cite{Maldacena} to leading order in large
$N$. When Super-Yang-Mills theory is coupled to a non-dynamical,
external metric, $g_{ij}$, the Weyl anomaly, ${\cal A}$, is the
response of the logarithm of the partition function, $F$, to a
scale transformation of that metric: $\delta F=\int d^4x\,{\sqrt
g} \delta\sigma {\cal A}$ when $\delta g_{ij}=2\delta\sigma
g_{ij}$. On general grounds ${\cal A}=a \,E+c\,I$ where $E$ is the
Euler density, $(R^{ijkl}R_{ijkl}-4R^{ij}R_{ij}+R^{2})/64$, and
$I$ is the square of the Weyl tensor,
$I=(-R^{ijkl}R_{ijkl}+2R^{ij}R_{ij}-R^{2}/3)/64$. A one-loop
calculation \cite{Duff1} gives ${\cal A}$ as the sum of
contributions from the six scalars, two fermions and gauge vector
of the Super-Yang-Mills theory, (all in the adjoint with dimension
$N^2-1$) \be\label{cftaa}{\cal A}={(6s+2f+g_v)(N^2-1)\over
16\pi^2}\,.\ee When the heat-kernel coefficients $s$, $f$, and
$g_v$ are expressed in terms of $E$ and $I$ this becomes
\be\label{cfta}{\cal A}=-{(N^2-1)(E+I)\over \pi^2},\ee so
$a=c=-(N^2-1)/(2\pi^{2})$ and supersymmetry protects this from
higher-loop corrections. Henningson and Skenderis showed that the
tree-level calculation in the bulk reproduces the leading $N^2$
piece by solving the Einstein equations perturbatively near the
boundary. We would expect that the $-1$ piece is due to string
loops in the bulk that to this order can be approximated by field
theory loops, but these depend on much more than just classical
General Relativity, and reproducing them provides a more stringent
test of the Maldacena conjecture sensitive to the detailed
particle content of the bulk IIB Supergravity theory. In
\cite{testn} we showed that the bulk Supergravity one-loop
contributions to $a-c$ vanished when summed over each
supermultiplet confirming the conjecture. In this letter we will
complete this calculation of the Weyl anomaly by computing $a$
itself and showing that it does indeed reproduce the $-1$ piece.

The one-loop contribution to $\cal A$ from bulk fields was found
in \cite{us} using Schr\"odinger functional methods that are
particularly appropriate to the AdS/CFT correspondence because,
being Hamiltonian, they apply four-dimensional technology to the
study of fields on a five-dimensional manifold with a boundary.
The result can be expressed \cite{us3} as \be \delta {\cal
A}=-\sum {(\Delta-2)a_{2}\over 32\pi^{2}} \label{asum} \ee where
the sum is taken over all the fields in IIB Supergravity
compactified on $AdS_{5}\times S^{5}$, $\Delta$ is the scaling
dimension of the associated boundary operator, and $a_{2}$ is a
four-dimensional heat-kernel coefficient (multiplied by $-1$ for
anti-commuting fields). Deriving this requires decomposing the
five-dimensional components of fields into those appropriate to
the four-dimensional boundary.

In deriving (\ref{asum}) the $AdS$ metric was taken to be

\be ds^2={1\over t^2}\left( l^2\,dt^2+\sum_{i,j}\hat
g_{ij}\,dx^i\,dx^j\right),\qquad t>0 \label{newmet} \ee which
satisfies the Einstein equations with cosmological constant
$-6/l^{2}$ provided $\hat g_{ij}$, (which is proportional to the
boundary metric), is Ricci flat. In this case $E=-I$ so that $\cal
A$ is proportional to $a-c$. To find $a$ itself it is convenient
to take a constant curvature boundary for which
$R_{ijkl}=(g_{ik}g_{jl}-g_{il}g_{jk})R/12$, $R_{ij}=Rg_{ij}/4$,
$I=0$ and $E= R^2/384$. The solution to Einstein's equations is
obtained by multiplying $\hat g_{{ij}}$ in (\ref{newmet}) by
$(1-\hat R t^{2}l^{2}/48)^{2}$, where $\hat R$ is the curvature
constructed from $\hat g_{ij}$. The effect of this extra piece on
the decomposition of five-dimensional fields into four-dimensional
variables is to introduce into the four-dimensional operators
precisely those couplings to $\hat R$ that render them conformally
covariant. Thus $a_2$ for a five-dimensional gauge field is the
heat-kernel coefficient for the operator associated with a
four-dimensional gauge field, whilst that for a minimally coupled
five-dimensional scalar is associated with a conformally coupled
four-dimensional scalar.

The scaling dimensions $\Delta$ are related to the bulk masses
which were originally worked out in \cite{Kim}. In Table 1 we
display the corresponding values of $\Delta-2$. The multiplets are
labelled by an integer $p\ge 2$, and the fields form
representations of $SU(4)\sim SO(6)$. The four-dimensional
heat-kernel coefficients have also been known for a long time and
we use the values given by \cite{Duff2,Barv}. In Table 2 we list
these for the cases of a Ricci flat boundary.

If we denote the values of $a_{2}$ for the fields $\phi$, $\psi$,
$A_\mu$, $A_{\mu\nu}$, $\psi_\mu$, $h_{\mu\nu}$ by $ s,f,v,a,r, $
and $g$ respectively then the contribution from a generic ($p\ge
4$ ) multiplet is \bq \left(\sum (\Delta-2)a_2\right)_{p\ge 4}&=&
(-4s+4a+r+f+2v){p\over 3}+\nonumber\\
(-105s-g-26a-8r-72f-48v){p^{3}\over 12}
&+&(16v+20f+10a+4r+25s+g){p^{5}\over 12} \eq whilst for the $p=3$
multiplet it is \be \left(\sum (\Delta-2)a_2\right)_{p=3}=
244f+18g+266s+218v+148a+64r\,. \ee The $p=2$ multiplet contains
gauge fields requiring the introduction of Faddeev-Popov ghosts.
Their parameters are given in Table 3 along with the decomposition
of the five-dimensional components of fields into four-dimensional
pieces. \be 12 v- 30  s +6 r-10f +2g \ee and if we include the
scalars, spinors and antisymmetric tensors the total contribution
of the $p=2$ multiplet is \be \left(\sum
(\Delta-2)a_2\right)_{p=2}=12 v- 6  s +6 r+6f +2g +12a \ee
Substituting the values of the heat kernel coefficients for a
Ricci flat boundary shows that the contribution of each
supermultiplet vanishes implying that $a=c$ \cite{testn}. However
if we do not specialise to this case we have to deal with the sum
over multiplets labelled by $p$. We will evaluate this divergent
sum by weighting the contribution of each supermultiplet by $z^p$.
The sum can be performed for $|z|<1$, and we take the result to be
a regularisation of the weighted sum for all values of $z$.
Multiplying this by $1/(z-1)$ and integrating around the pole at
$z=1$ gives a regularisation of the original divergent sum. This
yields \be \sum (\Delta-2)a_2=8s+4f+2v \ee which remarkably
depends only on the heat-kernel coefficients of fields in the
Super-Yang-Mills theory. By decomposing a five-dimensional vector
into longitudinal and transverse pieces and solving the
Schr\"odinger equation for them, it can be seen that the
heat-kernel coefficient for a vector field, $v$, is related to
that for the four-dimensional (gauge-fixed) Maxwell operator,
$v_0$, as $v = v_0 + 2s - 2s_0$ where $s_0$ is the coefficient for
a minimally coupled four-dimensional scalar (Faddeev-Popov ghost),
showing $v-2s = v_0 -2s_0 = g_v$ \cite{mnu-prep}. Therefore we
finally arrive at the one-loop contribution to the Weyl anomaly
\be \delta {\cal A}=-\sum {(\Delta-2)a_{2}\over
32\pi^{2}}=-{6s+2f+g_v\over 16\pi^2} \label{afsum} \ee which is
precisely what is needed to reproduce the subleading term in the
exact Weyl anomaly of Super-Yang-Mills theory and verify the
Maldacena conjecture.

It is worth emphasising that  $a$ received non-trivial
contributions from all the supermultiplets, not just the $p=2$
multiplet containing gauge fields, in contrast to \cite{chu}. This
indicates that although bulk tree-level solutions might be
constructed by a `consistent' truncation of the full IIB
Supergravity to this single multiplet, as in studies based on
gauged ${\cal N}=8$ Supergravity, such a procedure would miss loop
effects in the bulk that contribute to the Super-Yang-Mills theory
at sub-leading order. So, for example, the application of
(\ref{asum}) to the spectrum of \cite{warner} fails to produce the
expected subleading correction to the coefficient $c$ for the
infra-red fixed point of the RG flow driven by adding certain mass
terms to the ${\cal N}=4$ Super-Yang-Mills theory to break the
supersymmetry down to ${\cal N}=1$.

\begin{table}[b]
\begin{center}
\caption{Mass spectrum. The supermultiplets (irreps of U(2,2/4))
are labelled
   by the integer $p$. Note that the doubleton ($p=1$) does not appear in
the
   spectrum. The $(a,b,c)$ representation of $SU(4)$ has dimension
   $(a+1)(b+1)(c+1)(a+b+2)(b+c+2)(a+b+c+3)/12$, and a subscript $c$
indicates
   that the representation is complex. (Spinors are four component Dirac
   spinors in $AdS_5$).}
\label{spec} \vskip .3cm
  \begin{tabular}{|cccc|}
\hline

     Field  & $SO(4)$ rep$^{\rm n}$ & $SU(4)$ rep$^{\rm n}$ &
$\Delta-2$      \\

  \hline

$\phi^{(1)}$ & $(0,0)$ & $(0,p,0)$ & $p-2$,\quad $p\ge2$ \\
$\psi^{(1)}$ & $(\hf,0)$ & $(0,p-1,1)_c$ & $p-3/2$,\quad $p\ge2$ \\
$A_{\mu\nu}^{(1)}$ & $(1,0)$ & $(0,p-1,0)_c$ & $p-1$,\quad
$p\ge2$ \\
\hline $\phi^{(2)}$ & $(0,0)$ & $(0,p-2,2)_c$ & $p-1$,\quad
$p\ge2$
\\
$\phi^{(3)}$ & $(0,0)$ & $(0,p-2,0)_c$ & $p$,\quad $p\ge2$
\\
$\psi^{(2)}$ & $(\hf,0)$ & $(0,p-2,1)_c$ & $p-1/2$,\quad $p\ge2$ \\
$A_\mu^{(1)}$ & $(\hf,\hf)$ & $(1,p-2,1)$ & $p-1$,\quad $p\ge2$
\\
$\psi_\mu^{(1)}$ & $(1,\hf)$ & $(1,p-2,0)_c$ & $p-1/2$,\quad
$p\ge2$
\\
$h_{\mu\nu}$ & $(1,1)$ & $(0,p-2,0)$ & $p$,\quad $p\ge2$ \\

\hline
$\psi^{(3)}$ & $(\hf,0)$ & $(2,p-3,1)_c$ & $p-1/2$,\quad $p\ge3$ \\
$\psi^{(4)}$ & $(\hf,0)$ & $(0,p-3,1)_c$ & $p+1/2$,\quad $p\ge3$ \\
$A_\mu^{(2)}$ & $(\hf,\hf)$ & $(1,p-3,1)_c$ & $p$,\quad
$p\ge3$ \\
$A_{\mu\nu}^{(2)}$ & $(1,0)$ & $(2,p-3,0)_c$ & $p$,\quad $p\ge3$
\\
$A_{\mu\nu}^{(3)}$ & $(1,0)$ & $(0,p-3,0)_c$ & $p+1$,\quad
$p\ge3$ \\
$\psi_\mu^{(2)}$ & $(1,\hf)$ & $(1,p-3,0)_c$ & $p+1/2$,\quad
$p\ge3$
\\
\hline
$\phi^{(4)}$ & $(0,0)$ & $(2,p-4,2)$ & $p$,\quad $p\ge4$ \\

$\phi^{(5)}$ & $(0,0)$ & $(0,p-4,2)_c$ & $p+1$,\quad $p\ge4$
\\
$\phi^{(6)}$ & $(0,0)$ & $(0,p-4,0)$ & $p+2$,\quad $p\ge4$ \\
$\psi^{(5)}$ & $(\hf,0)$ & $(2,p-4,1)_c$ & $p+1/2$,\quad $p\ge4$ \\
$\psi^{(6)}$ & $(\hf,0)$ & $(0,p-4,1)_c$ & $p+3/2$,\quad $p\ge4$ \\
$A_\mu^{(3)}$ & $(\hf,\hf)$ & $(1,p-4,1)$ & $p+1$,\quad $p\ge4$
\\
\hline
  \end{tabular}
  \end{center}
\end{table}

\begin{table}[tbp]
\begin{center}
\caption{Anomaly coefficients of massive fields on $AdS_5$. Note
that the massive vector coefficient is $v_0+2s-2s_0$ where
$v_0,s,s_0$ are respectively, the coefficients for the 4d
gauge-fixed Maxwell operator, a conformally coupled scalar, and a
minimally coupled scalar. } \label{coeffs} \vskip .3cm
\begin{tabular}{|ccc|}
\hline
Field & $R_{{ij}}=0$: & Constant $R$:\\
        & $180 a_{2}/R_{ijkl}R^{ijkl}$ & $180a_2/R^2$ \\
\hline
$\phi$ & 1 & -1/12\\
$\psi$  & 7/2 & -11/12\\
$A_\mu$ & -11 & 29/3\\
$A_{\mu\nu}$&  33 & 19/4\\
$\psi_\mu$ & -219/2 & -61/4\\
$h_{\mu\nu}$ & 189 &747/4\\
\hline

   \end{tabular}
  \end{center}
\end{table}

\begin{table}[t]
\begin{center}
\caption{Decomposition of gauge fields for the massless
multiplet.} \label{ghosts} \vskip .3cm
\begin{tabular}{|c|cccc|}
\hline
Original field & Gauge fixed fields & $\Delta-2$ & $R_{{ij}}=0$:& Constant $R$:\\
    &  &  & $180 a_{2}/R_{ijkl}R^{ijkl}$ & $180a_2/R^2$\\
\hline
$A_\mu$ & $A_i$ & 1 & -11 & 29/3\\
({\bf 15} of $SU(4)$)       & $A_0$ & 2 & 1 & -1/12\\
         & $b_{FP}$, $c_{FP}$ & 2 & -1 & 1/12\\
\hline
$\psi_\mu$ & $\psi_i^{\rm irr}$ & 3/2 & -219/2 & -61/4\\
&            $\gamma^i\psi_i$ & 5/2 & 7/2 & -11/12\\
({\bf 4} of $SU(4)$)  & $\psi_0$ & 5/2 & 7/2 & -11/12\\
& $\lambda_{FP}$, $\rho_{FP}$ & 5/2 & -7/2 & 11/12\\
& $\sigma_{GF}$ & 5/2 & -7/2 & 11/12\\
\hline
$h_{\mu\nu}$ & $h_{ij}^{\rm irr}$ & 2 & 189 & 727/4\\
($SU(4)$ singlet) & $h_{0i}$ & 3 & -11 & 29/3\\
& $h_{00}$, $h_\mu^\mu$ & $\sqrt{12}$& 1 & -1/12\\
& $B^{FP}_0$,$C^{FP}_0$ &$\sqrt{12}$& -1 & 1/12\\
& $B^{FP}_i$,$C^{FP}_i$ & 3 & 11 & -29/3\\
\hline
   \end{tabular}
  \end{center}
\end{table}


\begin{thebibliography}{88}
\bibitem{Henningson} M. Henningson and K. Skenderis JHEP 9807 (1998),
023.

\bibitem{Maldacena} J. Maldacena, Adv.Theor.Math.Phys.2 (1998), 231.

\bibitem{Duff1} M.J. Duff, Class. Quant. Grav. 11 (1994) 1387, and refs. therein.

\bibitem{testn} P. Mansfield and D. Nolland Phys.Lett.B495:435-439,
(2000).

\bibitem{us}
P. Mansfield and D. Nolland, JHEP 9907 (1999), 028.

\bibitem{us3}
P. Mansfield and D. Nolland, Phys.Lett.B515:192-196,2001.

\bibitem{Kim}
H.J.~Kim, L.J.~Romans and P.~van Nieuwenhuizen, Phys. Rev. D32
(1985), 389.

\bibitem{Duff2} S.M. Christensen and M.J. Duff, Nucl.Phys.B 154
(1979)301.

\bibitem{Barv} A.O. Barvinsky and G.A. Vilkovisky, Phys. Rep.
119(1985)1.

\bibitem{mnu-prep} P. Mansfield, D. Nolland and T. Ueno, in
preparation.

\bibitem{chu} A. Bilal and C-S. Chu, Nucl. Phys. B562 (1999), 181.

\bibitem{warner} D.Z. Freedman, S.S. Gubser, K. Pilch, N.P. Warner,
Adv.Theor.Math.Phys. 3 (1999) 363-417.
\end{thebibliography}
\end{document}